\documentstyle[12pt]{article}

\begin{document}

\baselineskip=6.8mm

\newcommand{\TeV}{\,{\rm TeV}}
\newcommand{\GeV}{\,{\rm GeV}}
\newcommand{\MeV}{\,{\rm MeV}}
\newcommand{\keV}{\,{\rm keV}}
\newcommand{\eV}{\,{\rm eV}}
\newcommand{\Tr}{{\rm Tr}\!}
\renewcommand{\arraystretch}{1.2}
\newcommand{\be}{\begin{equation}}
\newcommand{\ee}{\end{equation}}
\newcommand{\bea}{\begin{eqnarray}}
\newcommand{\eea}{\end{eqnarray}}
\newcommand{\ba}{\begin{array}}
\newcommand{\ea}{\end{array}}
\newcommand{\bmat}{\left(\ba}
\newcommand{\emat}{\ea\right)}
\newcommand{\refs}[1]{(\ref{#1})}
\newcommand{\ler}{\stackrel{\scriptstyle <}{\scriptstyle\sim}}
\newcommand{\ger}{\stackrel{\scriptstyle >}{\scriptstyle\sim}}
\newcommand{\lag}{\langle}
\newcommand{\rag}{\rangle}
\newcommand{\ns}{\normalsize}
\newcommand{\cm}{{\cal M}}
\newcommand{\gr}{m_{3/2}}
\newcommand{\p}{\partial}
\def\tl{{\tilde{l}}}
\def\tL{{\tilde{L}}}
\def\bd{{\overline{d}}}
\def\tL{{\tilde{L}}}
\def\a{\alpha}
\def\b{\beta}
\def\g{\gamma}
\def\c{\chi}
\def\d{\delta}
\def\D{\Delta}
\def\db{{\overline{\delta}}}
\def\Db{{\overline{\Delta}}}
\def\e{\epsilon}
\def\l{\lambda}
\def\n{\nu}
\def\m{\mu}
\def\nt{{\tilde{\nu}}}
\def\p{\phi}
\def\P{\Phi}
\def\x{\xi}
\def\r{\rho}
\def\s{\sigma}
\def\t{\tau}
\def\th{\theta}
\renewcommand{\Huge}{\Large}
\renewcommand{\LARGE}{\Large}
\renewcommand{\Large}{\large}

\begin{titlepage}
\title{\bf Models of Light Singlet Fermion and Neutrino Phenomenology \\
                          \vspace{-4cm}
                          \hfill{\ns IC/95/76\\}
                          \hfill{\ns PRL-TH/95-7\\}
                          \hfill{\ns hep-ph/9505275\\[.3cm]}
                          \hfill{\ns May 1995}
                          \vspace{3cm} }

\author{ E.~J.~Chun$^\dagger$ \hspace{.4cm}
         Anjan S.~Joshipura$^*$ \hspace{.4cm}
         A.~Yu.~Smirnov$^{\dagger,\#}$ \\[.5cm]
  {\ns\it $^\dagger$International Center for Theoretical Physics}\\
  {\ns\it P.~O.~Box 586, 34100 Trieste, Italy} \\[.3cm]
  {\ns\it $^*$Theoretical Physics Group, Physical Research Laboratory}\\
  {\ns\it Navarangpura, Ahmedabad, 380 009, India} \\[.3cm]
  {\ns\it $^\#$ Institute for Nuclear Research, Russian Academy of Sciences}\\
  {\ns\it 117312 Moscow, Russia} }
\date{}
\maketitle
\vspace{2cm}
\begin{abstract} \baselineskip=7mm
{\ns We suggest that a singlet fermion $S$ exists beyond the standard
see-saw structure.  It mixes with light neutrinos via
interactions with the right-handed neutrino components, so that $\n_e \to S$
conversion solves the solar neutrino problem.  Supersymmetry endowed with
R-symmetry is shown to give a natural framework for existence, mass scale
and mixing ($\sin^22\th_{es} \sim (0.1-1.5)\cdot 10^{-2}$) of such a fermion.
Models with an approximate horizontal symmetry are constructed, which embed
the fermion $S$ and explain simultaneously solar, atmospheric, hot dark matter
problems as well as may predict the oscillation $\bar{\n}_\m \to \bar{\n}_e$
in the region of sensitivity of KARMEN and LSND experiments.}
\end{abstract}

\thispagestyle{empty}
\end{titlepage}

\section{Introduction}

The solar neutrino problem \cite{solar}, the deficit of muon neutrinos in
atmospheric neutrino flux \cite{atmos},  the large scale structure of
the Universe \cite{hdm} and possible candidate events in a search for
$\bar{\n_\m} \to \bar{\n_e}$ oscillations \cite{LSND} (see however
\cite{hill}) give indications on non-zero neutrino masses and lepton
mixing.  Simultaneous explanation of all (or some)
of these problems may call for the existence of more than three light
neutrinos which mix among themselves \cite{calmo}.
Strong bounds on the number of neutrino
species both from the invisible $Z^0$--width and from primordial
nucleosynthesis (NS) \cite{bbn} require the additional neutrino
to be sterile (singlets of $SU(2) \times U(1)$).
The right-handed (RH) components of known neutrinos are natural candidates
for such sterile states. However, in such a case one has to depart from
the conventional see-saw mechanism which implies
large masses to the RH components.

A number of schemes with light sterile neutrinos has been suggested
\cite{valle1}--\cite{anjan}.
Most of them are based on radiative mechanism of mass generation  or on
some hybrid schemes which include both the elements of the see-saw and
radiative mechanisms. In these schemes sterile neutrino is considered on
the same footing as the usual neutrinos.  The lepton number is broken
typically at the electroweak scale.

\bigskip

We will consider another possibility.  We suggest that usual
see-saw mechanism works with all three right-handed neutrinos
having large Majorana masses: $M \sim 10^{10}-10^{12}$ GeV.
At the same time the theory contains an additional singlet fermion $S$
which has its origin beyond the standard lepton structure.
The singlet $S$ is very light and mixes with neutrinos.

Supersymmetry (SUSY) can provide a natural justification for the existence
of $S$.  Many extensions of the standard model contain singlet scalar fields:
singlet majoron \cite{majoron}, invisible axion \cite{axion}, or
scalars for spontaneous generation of the $\m$--term \cite{mu}, etc..
The supersymmetric partners of such scalars could be identified with $S$.
Moreover, SUSY can play a crucial role in the determination of mass scales
in the singlet sector.

In this paper we consider possible origin of light fermion $S$, its
mass and mixing with light neutrinos.  The models with $S$ are constructed
so that they can simultaneously explain the above mentioned neutrino
anomalies.

\section{Light singlet fermion and the solar neutrino problem}

Primordial nucleosynthesis (as well as the data from SN87A) gives strong
bound on the oscillation of active neutrinos into sterile neutrino
\cite{nsbound}.
This practically excludes $\n_\m \to S$ oscillations as a
solution of the atmospheric neutrino problem.
The singlet fermion with mass in the eV--range could be considered as a
candidate for hot dark matter (HDM) \cite{calmo}.  However, if its density
satisfies the NS bound on the number of additional neutrino species:
$\d N_\n \ler 0.1$, it can not reproduce the optimal parameters \cite{hdm}
for the  large scale structure formation in the Universe: $m_S \sim
(2-5) \eV$ and $\Omega_s \simeq 0.2$, where $\Omega_s$ is the energy density
of $S$ in the Universe in the unit of the critical density.
Therefore it may happen that the only place where singlet fermion
plays a role is the solar neutrino problem.

Let us find the region of parameters $\D m^2$ and $\sin^22\th_{es}$,
where $\th_{es}$ is the mixing angle of $\n_e$ with $S$
for which the resonance conversion $\n_e \to S$ inside the Sun can
explain the existing data.
It is instructive to compare the sterile ($\n_e \to S$) and the active
neutrino ($\n_e \to \n_f$) cases. The $\n_e \to S$ solution of the solar
neutrino problem differs from the $\n_e \to \n_\m\,(\n_\t)$ solution in
two ways.

(1) The effective density, $\r_s$, for $\n_e \to S$ conversion
is smaller than that, $\r_f$, for $\n_e \to \n_\m$ conversion:
$$ {\r_s \over \r_f} = {Y_e - {1\over2}Y_n \over Y_e}\;. $$
Here $Y_e$  and $Y_n$ are the number densities of the electron and the
neutron per nucleon, respectively.
In the center of the Sun one gets  $\r^c_s/\r^c_f \simeq 0.76$ \cite{smir}.
The central density $\r^c$ determines  position of the adiabatic edge of the
suppression pit: $\left( E/\D m^2 \right)_a \propto 1/\r^c$.  Consequently,
in the $\n_e \to S$ case the adiabatic edge is shifted to larger $E/\D m^2$ in
comparison with the flavour case: $\left( E/\D m^2 \right)_s = \left(E/\D
m^2\right)_f \cdot \r_f/\r_s$.  The position of the nonadiabatic edge depends
on $\dot{\r}/\r$ and the difference between  the flavour
and sterile cases is practically negligible.

In the region of small mixing solutions \footnote{Large mixing domain is
excluded by primordial nucleosynthesis  data}, the allowed values of $\D m^2$
are determined essentially by $\left(E/\D m^2\right)_a$  and by Gallium
experiment data \cite{krasmir}.  Therefore the shift of the adiabatic edge
for $\n_e \to S$ to larger $E/\D m^2$ results in  corresponding shift of
$\D m^2$ to smaller values:
\be \label{shift}
 \D m^2|_s \simeq {\r^c_s \over \r^c_f}\,\D m^2|_f \sim 0.76\, \D m^2|_f \;.
\ee

(2) The fermion $S$ has no weak interactions, and therefore $S$ flux from
$\n_e \to S$ conversion does not contribute to the Kamiokande signal
($\n e \to \n e$ scattering) in contrast with
flavour case, where $\n_\m$ interacts via neutral currents.  This influences
the allowed region of mixing angles.  Indeed, for unfixed original Boron
neutrino flux (which has the largest, $\sim 50 \%$, theoretical
uncertainties) the bound on $\sin^22\th_{es}$ is determined by
the ``double ratio'' \cite{krasmir}:
$$ R_{H/K} \equiv {R_{Ar} \over R_{\n_e}}\;, $$
where $R_{Ar} \equiv Q^{obs}_{Ar}/Q^{SSM}_{Ar}$ and
$R_{\n_e} \equiv \P^{obs}_{B}/\P^{SSM}_{B}$ are the suppressions of signals
in $Cl$--$Ar$ and Kamiokande experiments, respectively. Here
$Q^{SSM}_{Ar}$, $\P^{SSM}_{B}$ are the predictions in the reference model
(e.g. \cite{bp}) and $Q^{obs}_{Ar}$, $\P^{obs}_{B}$ are the observable signals.
Due to the $\n_\m$ ($\n_\t$)--effect, $R^s_{\n_e}$ in the sterile case is
smaller than  $R^f_{\n_e}$ in the flavour case, and since
$R_{Ar}$ is the same in both cases one gets $R^s_{H/K} > R^f_{H/K}$.
With diminishing of $\th_{es}$, the suppression of $\P_B$ due to conversion
weakens and the effect of $\n_\m$ ($\n_\t$) decreases, therefore
$R^s_{H/K}$ approaches $R^f_{H/K}$.   As a consequence, the lower
bound on $\th_{es}$ coincides practically with that for flavour conversion:
$\sin^22\th_{es} \ger (0.8-1.0)\cdot 10^{-3}$ \cite{krasmir}.
On the contrary, with increase of $\th_{es}$ the suppression of
$\P_B$ due to the conversion becomes stronger, so that  $R_{Ar} \to 0$ and
$R^s_{\n_e} \to 0$; at the same time due to the neutral current effect of
$\n_\m (\n_\t)$, $R^f_{\n_e} \to 0.16$.
Therefore $R^f_{H/K} \to 0$, whereas $R^s_{H/K}$ does not change
strongly.  We have found $R^s_{H/K} \simeq$ 0.77,  0.74, 0.72,  0.69 for
$\sin^22\th_{es} \simeq$  $2\cdot10^{-3}$,  $5\cdot10^{-3}$, $10^{-2}$,
$2\cdot10^{-2}$, respectively.
The experimental value of the double ratio is $R_{H/K} = 0.67 \pm 0.11$.
However for large  $\sin^22\th_{es}$ the original flux of Boron neutrinos
should be large (to  compensate for strong suppression effect).  If we
restrict $\P_B \leq  1.5\P^{SSM}_B$, then the bound on the mixing angle
becomes: $\sin^22\th_{es} \ler 1.5\cdot 10^{-2}$. This also satisfies the NS
bound \cite{nsbound}.

Resonance conversion implies that $m_S > m_{\n_e}$ and if there is no
fine-tunning of masses, $m_S \simeq \sqrt{\D m^2}$.
Thus using \refs{shift} and known results for flavour conversion  as well as
bounds on $\sin^22\th_{es}$ discussed above we get the following range of the
parameters:
\bea \label{parameters}
 m_S &\simeq& (2-3)\cdot 10^{-3} \eV \nonumber\\
 \sin\th_{es} &\simeq& \tan\th_{es} \simeq (2-6)\cdot 10^{-2}  \;.
\eea

\section{Mass and mixing of singlet fermion via right-handed neutrino}

Let us consider the following Lagrangian,
\be \label{base}
 {\cal L} = {m_e \over \lag H_2 \rag}L_e\n_e^cH_2 + {M_e\over 2}\n^c_e\n^c_e
              + m_{es}\n^c_e S \;,
\ee
where $L_e$ is the lepton doublet, $H_2$ is the Higgs doublet and $\n_e^c$ is
the right-handed neutrino component.
We suggest that there is no direct coupling of $S$ with $L_e$ due to a certain
symmetry, and the mass term $SS$ is absent or negligibly small.
The Dirac mass $m_e$ and the mixing mass $m_{es}$
are much smaller than the Majorana mass $M_e$: $m_e, m_{es} << M_e$.
The Lagrangian \refs{base} leads to the mass matrix in the basis
$(S, \n_e, \n^c_e)$:
\be \label{mm1}
 {\cal M} = \bmat{ccc} 0 & 0 & m_{es}\\ 0 & 0 & m_e \\ m_{es} & m_e & M_e
              \emat \;.
\ee
The diagonalization of \refs{mm1} is straightforward: one combination of
the $\n_e$ and $S$,
$$ \n_0 = \cos\th_{es}\,\n_e + \sin\th_{es}\, S\;, \nonumber $$
is massless, and the orthogonal combination,
$$ \n_1 = \cos\th_{es}\, S - \sin\th_{es}\, \n_e \;, \nonumber $$
acquires a mass via the see-saw mechanism:
\be \label{m1}
 m_1 \simeq -{m_e^2 + m_{es}^2 \over M_e}\;.
\ee
The mass of the heavy neutrino is $\simeq M_e$.
The $\n_e$--$S$ mixing angle is determined by
\be \label{th}
 \tan\th_{es} = {m_e \over m_{es}} \;,
\ee
and correspondingly $\sin^22\th_{es} = 4[m_{es}m_e / m_{es}^2 + m_e^2]^2$.\
Taking for $m_e$ the typical Dirac mass of the first generation:
$m_e \sim (1-5) \MeV$, and suggesting that $\n_e \rightarrow
S$ conversion explains the solar neutrino problem with $m_1 =m_S$
as in \refs{parameters}, we find
\be \label{mis}
 m_{es} = {m_e \over \tan\th_{es}} \simeq (0.02-0.3) \GeV \;.
\ee
According to \refs{m1} the RH mass scale is
\be \label{Mis}
 M_e \simeq m_{es}^2/m_1 = {m^2_e \over m_1 \tan^2\th_{es}}
   \simeq (10^8-3\cdot10^{10})\GeV\;.
\ee

\bigskip

Consider now the models which lead to the Lagrangian \refs{base} with
parameters \refs{mis} and \refs{Mis}.  The simplest possibility is to use
the $U(1)$ symmetry of lepton number and to
generate the masses in \refs{base} by VEV $\lag \s \rag$ of the scalar
singlet, $\s$.  Prescription of the lepton charges $(1,-1,-3,2) $ for
$(\n_e, \n^c_e, S, \s)$
admits the following interactions in the singlet sector:
\be \label{model1}
 {\cal L} = h \n^c_e \n^c_e \s + h' \n^c_e S {\s^2 \over M_{Pl}}
            + h'' SS{\s^3 \over M_{Pl}^2} \;,
\ee
where $M_{Pl}$ is the Planck mass.  The Lagrangian \refs{model1}
reproduces the mass terms of \refs{base} with $M_e=h\lag \s \rag$ and $m_{es}
\simeq h' \lag \s \rag^2/M_{Pl}$. The desired values of $M_e$ and $m_{es}$
\refs{mis} \refs{Mis} can be achieved with e.g.,
$\lag \s \rag \simeq 10^{10}$ GeV, $h\simeq 1$ and $h' \simeq 10^{-2}$.
The last term in \refs{model1} generates the Majorana
mass of $S$, $m_{SS} = h''\lag\s\rag^3/M_{Pl}$,  so that
all the neutrinos are massive. For $h'' \ler 10^{-4}$ one gets $m_1
\simeq m_{es}^2/M_e$ as before,  whereas the smallest mass is
$\simeq m_e^2 m_{SS}/m_{es}^2 < m_1$.

\bigskip

Let us consider the possible role of supersymmetry in the appearance of
the singlet fermion and in the determination of its properties.
In principle, $S$ can be a superpartner of the goldstone boson which appears
as a result of spontaneous violation of a certain global symmetry like
lepton number or Peccei-Quinn symmetry.  In this connection,
let us consider a SUSY model with spontaneous violation of lepton number.
The superpotential of the singlet majoron model is
\be \label{model2}
 W = {m_e \over \lag H_2\rag}L_e\n_e^cH_2 + f\n_e^c\n_e^c\s  - \l(\s\s'-M^2)y
   \;,
\ee
where lepton numbers of the superfields $(L_e,\n_e^c,\s,\s',H_2,y)$ are
$(1,-1,2,-2,0,0)$.  Lepton number is spontaneously broken by non-zero VEV's
of $\s$ and $\s'$.  As the result, the majoron and its fermionic partner,
the majorino, are massless in the supersymmetric limit.

The identification of the majorino with $S$ requires, however, the following
complication of the model.

(1) Supersymmetry breaking results in appearance of non-zero VEV of
$y$ which generates the mass of the majorino $S$: $m_{SS} = \l \lag y \rag$.
The soft-breaking terms $\l (A_y \s \s'- B_y M^2) y + \mbox{h.c.}$, where
$A_y, B_y \simeq {\cal O}(m_{3/2})$ are soft-breaking parameters
give \cite{gy}
\be \label{yvev}
 \lag y \rag \simeq {1\over 2\l} (A_y -B_y)\;,
\ee
and consequently too  big value of $m_{SS} \simeq  (A_y - B_y)/2 \sim
{\cal O}(m_{3/2})$,  whereas $A_y - B_y \ler 10^{-3}$ eV is needed.
One can get $A_y - B_y = 0$ at tree level in no-scale supergravity
or in the case of the non-minimal kinetic term discussed below.
However, non-zero value of $A_y - B_y$ will be
generated due to renormalization group evolution of soft-terms.
In order to suppress the mass below the solar neutrino mass scale
\refs{parameters}, tuning of parameters is needed: $f \ler 10^{-5}$, if all
three generation of leptons are taken into account.

(2) Mixing of neutrinos with the majorino implies violation of R-parity.
In \refs{model2} the mixing can be induced by the second term if sneutrino
$\tilde{\n}^c$ gets non-zero VEV $\lag \tilde{\n}_e^c \rag$.
The latter requires the introduction of terms like $\n_e^c F(X_i) + W'(X_i)$,
where $F(X_i)$ and $W'(X_i)$ are the functions of new superfields $X_i$.
They should be arranged in such a way that in the global SUSY limit  $F$ does
not get a VEV: $\lag F(X_i) \rag = 0$, and after lepton number breaking
linear term $\sim M^2 \n_e^c$ appears in the superpotential.  Then the
corresponding soft-term  will give the mixing mass.
One can find that the additional sector requires at least several new fields
with non-zero lepton numbers which leads to further complication.
R-parity violation is a general feature of models in which $S$ is
identified with fermionic superpartner of scalars acquiring non-zero VEV as
in models for majoron, axion and $\mu$--term.

\bigskip

The above problems can be avoided in models with R-parity conservation.
In this case, the lightest supersymmetric particle can be served as cold
dark matter of the Universe.
To preserve R-parity one should place the singlet $S$ in the superfield
with zero VEV. Consider the superpotential:
\be \label{model3}
 W = {m_e\over \lag H_2\rag} L_e\n_e^cH_2 + f\n_e^c\n_e^c\s + f'\n_e^c S y
        - {\l \over 2}(\s^2 - M^2)y \;.
\ee
Its structure is determined by the R--symmetry under which the fields carry
the R--charges:
$$ (1,1,-1,2,0,0) \quad \mbox{for} \quad (L_e,\n_e^c,S,y,\s,H_2) \;. $$
Note that the R--symmetry forbids the bare mass terms $SS$ as well as
the coupling $SS\s$.  Since lepton symmetry is explicitly broken
no majoron appears. In the global SUSY limit, $\s$ gets non-zero
VEV $\lag \s \rag \simeq M \sim 10^{11}$ GeV which generates the Majorana
mass of $\n_e^c$: $M_{e} = f \lag \s \rag$.

SUSY breaking induces the following soft-breaking terms in the scalar
potential:
\be \label{soft}
 V_{soft} =  \{ A_L {m_e \over \lag H_2 \rag} L_e\n_e^cH_2 +
    fA_\n \n_e^c\n_e^c\s + f' A_S \n_e^cSy - {\l \over2}(A_y\s^2 - B_yM^2)y
    + \mbox{h.c.} \} + \sum_i m_i^2 |z_i|^2 \;,
\ee
where $z_i$ denotes the fields appearing in the superpotential \refs{model3}
and $A_L$, etc., are the soft-breaking parameters.
Minimization of the potential shows the following:\\
(1) The fields $L_e, \n^c_e, S$ do not develop VEV and therefore R-parity is
unbroken. \\
(2) The field $y$ acquires non-zero VEV due to the soft-breaking terms
as in \refs{yvev}.
Consequently, the mixing mass for $S$ and $\n_e^c$ appears:
\be  \label{mis2}
m_{es}= {f'\over 2\l}(A_y-B_y)
\ee
Since $m_{es} >> m_1$, no strong tunning of $A_y-B_y$ is needed as in the
previous case \refs{model2}.
At $A_y-B_y \sim O(m_{3/2})$, the desired value of $m_{es}$ \refs{mis} can be
obtained by choosing $f'/\l \sim 10^{-3}-10^{-2}$.
However, more elegant possibility is that $A_y=B_y$ at the Planck scale
but a non-zero value  for $A_y - B_y$ is generated due to renormalization
group evolution through  the differences in interactions of $\s$ and $y$.
In this case  one expects
\be \label{mrad}
  m_{es} \sim {\bar{\l}^2\over 16\pi^2} m_{3/2} \;,
\ee
where $\bar{\l}$ represents a combination of the constants $\l,f$ and $f'$.
As a consequence, the value $m_{es}\sim 0.1$ GeV does not require
smallness of $\bar{\l}$ or $f'$.

The equality $A_y = B_y$ at the Planck scale can be achieved  by the
introduction of non-minimal kinetic term with mixings between the observable
and hidden sectors. Let us introduce the following K\"ahler potential:
\be
 K = C\overline{C} + C\overline{C}(a\frac{Z}{M_{Pl}} +
     \overline{a}\frac{\overline{Z}}{M_{Pl}}) + Z\overline{Z} \;,
\ee
where $C$ and $Z$ represent an observable and hidden sector field,
respectively.  Then usual assumption that the observable sector has no
direct coupling to the hidden sector in superpotential, $W=W(C) + W(Z)$,
leads to the universal soft-terms:
\be
 V_{soft} \sim m_{3/2} W(C) +  \mbox{h.c.} \;,
\ee
provided $\overline{a}= \lag W(Z) \rag / \lag M_{Pl} \partial W/\partial Z +
W(Z)\overline{Z}/M_{Pl} \rag $.
Note also that the field $C$ does not acquire a soft-breaking mass.
This mechanism can be generalized to arbitrary number of observable sector
fileds.  For our purpose  $C \equiv \s, y$, i.e., we couple $\s$ and $y$ to
the hidden sector field $Z$ with the above-mentioned choice for $a$.

Note that $\s$ field plays two-fold role in the model:
it gives Majorana mass of $\n^c$ and it also generates mixing of $\n^c$ with
$S$ by inducing a VEV for $y$  after the SUSY breakdown.
Moreover, $\s$ can be used to generate the $\m$--term via the
non-renormalizable interaction:
\be \label{muterm}
        {\s^2 \over M_{Pl}} H_1 H_2 \;.
\ee
The $\mu$--term can also be generated through the renormalizable interaction:
$y H_1 H_2$ in the case of $\lag y \rag \simeq {\cal O}(m_{3/2})$.

It is easy to incorporate the spontaneous violation of lepton number or/and
Peccei-Quinn symmetry into the model. As in \refs{model2} one should
introduce the superfield $\s'$ with lepton number $-2$ and zero R-charge
and replace the $\s^2$ term of \refs{model3} by $\s\s'$.
In this way the $\mu$--term \refs{muterm} can be naturally related to the
solution of the strong-CP problem via Peccei-Quinn mechanism \cite{muproblem},
and the majoron will coincide with the invisible axion \cite{langa}.

\section{Models with light singlet fermion}

Two other neutrinos, $\n_\m$ and $\n_\t$, can be included in the scheme
by adding to \refs{base} analogous terms with $L_\a$ and
$\n^c_\a$ ($\a = \m,\t$).  Then as in \refs{th},
mixing of these neutrinos with singlet $S$ is determined  by
$ \tan\th_{\a s} = m_{\a s}/m_\a$, where $m_{\a s}$ and $m_\a$ are the
corresponding mixing and Dirac masses.

Primordial nucleosynthesis gives strong bounds on the angles
$\th_{\a s}$ and/or on masses of light neutrino components: $
\sim m_\a^2/M_\a$.
Suppose that $S$ is family blind, and its couplings with
all neutrinos are universal: $ m_{e s} \simeq m_{\m s} \simeq m_{\t s}
\simeq (0.02-0.3)\,\GeV$.
Note that this mass scale (motivated by solar neutrinos) is of the order of
Dirac masses in the second generation.
Then with $m_{\m s} \sim m_\m \sim 0.3$
GeV, one gets $\tan\th_{\m s} \simeq 1$,  and if $m_2 \simeq m^2_\m/ M_\m
\simeq 0.1$ eV, the oscillation $\n_\m \rightarrow S$ could explain
the deficit of atmospheric neutrinos.  However, this possibility is strongly
disfavoured by NS data.
For $m_\m\simeq 1$ GeV one has $\sin^22\th_{\m s}\simeq (0.2-4)\cdot
10^{-2}$, and the NS bound \cite{nsbound} is satisfied if $\D m^2 \ler
(10^{-4}-10^{-3}) \eV^2$, or $m_2 \ler 3\cdot10^{-2}$ eV.
For the third generation ($m_\t \sim 100$ GeV), analogous figures are:
$\sin^22\th_{\t s} \simeq (0.2-5)\cdot 10^{-6}$ and $m_3 < 3$ eV.
Therefore, the cosmologically interesting masses of $\n_\t$ are admitted.
Note that the bound on $m_2$ form NS and values of $m_1$ and $m_3$ desired by
solar and HDM problems can be reproduces by moderate mass hierarchy of the RH
neutrinos: $M_\a \simeq 10^{10} - 10^{12}$ GeV.

\bigskip

To have simultaneously neutrinos as HDM and the solution of the atmospheric
neutrino problem via $\n_\m \rightarrow \n_\t$ oscillations one needs
$m_2 \simeq m_3 \simeq 2$ eV. In this case ($\D m^2 \simeq 4
\eV^2$) the NS bound: $\sin^22\th_{\m s} \ler 10^{-6}$
implies $m_{\m s}/m_\m < 5\cdot10^{-4}$ or $m_{\m s} \ler 0.5$ MeV  at
$m_\m \sim 1$ GeV, i.e. the coupling of $S$ with $\n_e$
should dominate: $m_{es}>>m_{\m s}$.

Both the dominance of $S$--$\n^c_e$ coupling and the near degeneracy
of neutrinos corresponding to the  second and the third generations
($m_2 \simeq m_3$) can arise as consequences of some family (horizontal)
symmetry.

Let us consider $U(1)^h$--symmetry with charge prescription
($0,1,-1$) for the first, the second and third generations of leptons,
respectively.   Each generation includes the left-handed doublet $L_\a$ and
the right-handed $\n_{\a R}, e_{\a R}$. Higgs doublets as well as new
particles $S,\s,\s',y$ have zero charges.  In the limit of exact symmetry,
the Higgs doublet and the singlet fermion $S$ can couple only with the
electron neutrino, reproducing the matrix \refs{mm1}.
The couplings for the second and third generations allowed by $U(1)^h$:
\be \label{model4}
  W = {m_\m \over \lag H_2 \rag} L_\m \n^c_\m H_2
      + {m_\t \over \lag H_2\rag} L_\t \n^c_\t H_2
      + {M\over \lag \s\rag} \n^c_\m \n^c_\t \s \;
\ee
lead to a mass matrix in $(\n_\m,\n_\t,\n^c_\m,\n^c_\t)$ basis
\be \label{mm2}
  \left( \ba{cccc} 0&0&m_\m&0\\ 0&0&0&m_\t\\ m_\m&0&0&M_{\m\t} \\
   0&m_\t&M_{\m\t}&0 \ea  \right) \;.
\ee
The mass matrix of charged leptons is diagonal.
The diagonalization of \refs{mm2} results in ZKM-type
(Zeldovich-Mahmoud-Konopinsky) light neutrino formed by $\simeq \n_\m$
and $\simeq \n_\t$ components with mass
\be \label{pseudm}
 m_2 = -m_3 = {m_\m m_\t \over M_{\m\t}} \;.
\ee
For $m_\m \sim 1$ GeV, $m_\t \sim 100$ GeV and $M_{\m\t} \sim 3\cdot 10^{10}$
GeV one gets $m_2 \simeq 3$ eV which is required for the HDM components.
In the limit of exact horizontal symmetry $\n_e$--$S$ and $\n_\m$--$\n_\t$
form two unmixed blocks and in particular, $m_{\m s} = m_{\t s} = 0$.

Family symmetry can be conserved at high scale but can be explicitly broken
by interactions with Higgs doublets.  Such breaking could be induced
spontaneously also by introducing  new Higgs doublets with non-zero
$U(1)^h$ charges ($\pm1$ or 2) or by non-renormalizable interactions of the
type: $L_e \n^c_\t H_2 \s_\m/M$, where $\s_\m$ has the charge +1 and acquire
the VEV at large scale, $\lag \s_\m \rag \sim 10^{-4} M$.

Violation of $U(1)^h$ leads to mass splitting in $\n_\m$--$\n_\t$ system as
well as to mixing between $\n_e$--$S$ and $\n_\m$--$\n_\t$ blocks.
Consider the phenomenological consequences of introducing $U(1)^h$ violation
separately in different sectors of the model.

(1) The non-diagonal Dirac mass terms $m_{\m\t} \n_\m \n^c_\t +
m_{\t\m} \n_\t \n^c_\m$ + h.c. result in mass-squared difference
\be \label{dmutau1}
 \D m^2_{23} \simeq  {4 m_{\t\m} m_2^2 \over m_\m } \;.
\ee
For the atmospheric neutrinos one needs $\D m^2_{23} \simeq 10^{-2} \eV^2$,
then for $m_2 \sim 2 \eV$ and  $m_\m \simeq 1$ GeV, it follows from
\refs{dmutau1} that $m_{\t\m}$ should be very small: $ \simeq (0.5-1)$
MeV. Mixing of $\n_\m$ and $\n_\t$ is practically maximal.

(2) The introduction of a diagonal element in the Majorana sector; e.g.,
$M_\t\n^c_\t\n^c_\t$, gives
\be \label{dmutau2}
 \D m^2_{23} \simeq 2{m_\m \over m_\t}
            \left(M_\t \over M_{\m\t} \right) m_2^2 \;,
\ee
and to have $\D m^2_{23} \simeq 10^{-2} \eV^2$ with $m_\m/m_\t \sim 2\cdot
10^{-2}$,  one needs $M_\t/M_{\m\t} \sim 0.1$.

(3) To get $\n_e$--$\n_\m$ mixing one can introduce the Dirac mass terms
$m_{e\m} \n_e\n^c_\m + m_{\m e} \n_\m\n^c_e$ + h.c..
Present sensitivity region of KARMEN and LSND: $\sin^22\th_{e\m} \sim
(3-5)\cdot10^{-3}$ corresponds to $m_{e\m}/m_\m
\simeq 3\cdot10^{-2}$, and consequently to $m_{e\m} \simeq 30$ MeV.
In this case $\n_\m$--$S$ mixing will also be generated with
$ \tan\th_{\m s} \sim (m_{es} m_{\m e})/(m_\m m_\t) \sim 3\cdot10^{-5}$
which is far below the NS bound.

(4) Violation of $U(1)^h$--symmetry implies in general a non-diagonal mass
matrix for the charged leptons.  In this case the lepton mixing matrix is
the product, $V=V_\n\cdot V_l^\dagger$, where $V_\n$ and $V_l$ diagonalize
the mass matrices of neutrinos and charged leptons, respectively.
Let us suppose for simplicity that the effects of $U(1)^h$ violation come
from $V_l$ only ($V_\n$ has two-block structure as before),
and moreover $V_l$ mixes essentially the first and the second generation
with the angle $\th_l$.  Then the oscillations $\n_e \leftrightarrow \n_\m$
are expected with $\D m^2 \simeq m^2_2$ and the depth $\sin^22\th_l$.
Also mixing between $S$ and $\n_\m$ appears, so that the $\n_\m \to S$
oscillations with $\D m^2 \simeq m^2_2$ will have the depth
$$ \sin^22\th_{s\m} \simeq \sin^22\th_{es}\cdot\sin^2\th_l \;. $$
For $\th_{es}$ and $\th_l$ fixed by solar neutrino data and the LSND/KARMEN
sensitivity, one finds $\sin^22\th_{s\m} \simeq (1-5)\cdot 10^{-6}$ which
can satisfy the NS bound. This model realizes the scenario described in
\cite{hdm}.

\bigskip

The $\n_\m$--$\n_\t$ mass splitting can be generated without explicit
$U(1)^h$ violation. The modified $U(1)^h$ charge prescription in the model
\refs{model3} with $\s'$: ($-1$,2,0,$-2$,2,1,0) for ($\n^c_e$,$\n^c_\m$,
$\n^c_\t$,$\s$,$\s'$,$S$,$y$) allows for the superpotential,
\bea \label{model5}
  W &=&\frac{m_e}{\lag H_2 \rag}\n_e\n_e^cH_2+
   \frac{m_\m}{\lag H_2 \rag}\n_{\m}\n_{\m}^cH_2+
    \frac{m_\t}{\lag H_2 \rag}\n_{\t}\n_{\t}^cH_2 \nonumber \\
   &+&\frac{M_e}{2 \lag \s'\rag}\nu_e^c\nu_e^c\s'
   +\frac{M_{\m\t}}{\lag \s \rag}\n_{\m}^c\n_{\t}^c\s+
   \frac{M_\t}{2}\n_{\t}^c\n_{\t}^c
   +\frac{m_{es}}{\lag y \rag}\n_e^cSy \;.
\eea
For $\n_e$--$S$ it reproduces the matrix \refs{mm1}, whereas for
$\n_\m$--$\n_\t$ system one gets the matrix \refs{mm2} with non-zero
$M_\t \n^c_\t \n^c_\t$ term, thus generating mass splitting \refs{dmutau2}.
However, the blocks $\n_e$--$S$ and $\n_\m$--$\n_\t$ remain decoupled,
and thus no observable effect in KARMEN/LSND is expected.

\section{Conclusion}

We suggest that a light singlet fermion $S$ whose existence is hinted by some
neutrino observations may have its origin beyond neutrino physics.  Such
a fermion can however be incorporated into the standard see-saw picture, where
interactions of $S$ with the heavy right-handed neutrinos can generate its
mixing with the light neutrinos. Such a mixing allows an understanding of
the lightness of $S$ without ad hoc introduction of very light scale.
The mixing mass parameter $m_{es}\simeq (0.02-0.3)$ GeV leads to the mass of
the singlet and its mixing with electron neutrino in the region $m_1 \simeq
(2-3)\cdot 10^{-3}$ eV and $\sin^22\th_{es} \simeq (1-15)\cdot 10^{-3}$,
where the $\n_e \to S$ resonance conversion gives a good fit of all solar
neutrino data.

\bigskip

Supersymmetry can  provide a framework within which the existence and the
desired properties of such a light fermion follow naturally.
There is a number of models with singlet scalars which acquire VEV and are
introduced to break symmetries such as  lepton number and Peccei-Quinn
symmetry, or to generate $\mu$--term, etc..
However, identifying $S$ with the fermionic superpartner of such scalars
implies violation of R-parity, and further complication of model.
We have considered a specific example with $S$ identified as the majorino.
It may be possible to suppress the mass of $S$ generated after SUSY breakdown
by introducing non-minimal K\"ahler potentials.

The conservation of R-parity requires for the fermion $S$ to be
a component of singlet superfield which has no VEV.
This allows to construct simple model \refs{model3}  in which the
properties (mass and mixing) of $S$ follow from the conservation of
R-symmetry.  The singlet field is mixed with RH neutrinos by the
interaction with the field $y$ which can acquire VEV radiatively after soft
SUSY breaking.  The model can naturally
incorporate the spontaneous violation of Peccei-Quinn symmetry or/and lepton
number.  The fields involved can spontaneously generate the $\m$--term.

\bigskip

Approximate horizontal (family) $U(1)^h$ symmetry as in \refs{model4}
provides simultaneous explanations for the predominant coupling of $S$ to
the first generation (thus satisfying the NS bound) and for the pseudo-Dirac
structure of $\n_\m$--$\n_\t$ needed in solving the atmospheric neutrino and
hot dark matter problem.
Breaking of $U(1)^h$ can be arranged in such a way that the
parameters of $\bar{\n}_\m \to \bar{\n}_e$ oscillations are in the region of
sensitivity of LSND and KARMEN experiments.

Future solar neutrino experiments will allow to prove or reject the
hypothesis of the $\n_e \to S$ conversion in the Sun \cite{bilenky}
and thus  to test the models elaborated in this paper.

\bigskip

{\it Note added}: When our work was practically accomplished we encountered
the paper \cite{maroy} discussing non-supersymmetric model based on discrete
symmetry in which sterile neutrino mixes with usual light neutrinos via RH
components.
Our results have been reported at XXX Rencontres de Moriond, March 11-18
(1995), Les-Arcs Savoie, France (to be published).

\bigskip

{\bf Acknowledgement:} A.S.J. wants to thank ICTP for its hospitality during
his visit.

\end{document}